\documentclass{iau}

\usepackage{amsmath}
\usepackage{graphics}
\usepackage{multirow}
\usepackage{hyperref}
\usepackage{graphicx}
\newcommand{\apj}{ApJ}
\newcommand{\apjl}{ApJL}
\newcommand{\apjs}{ApJS}
\newcommand{\mnras}{MNRAS}
\newcommand{\aap}{A\&A}
\newcommand{\aj}{AJ}
\newcommand{\araa}{ARA\&A}

\begin{document}

\lefttitle{Radio Outflows in RQ KSRs}
\righttitle{Salmoli Ghosh et al.}

\jnlPage{1}{7}
\jnlDoiYr{2025}
\doival{10.1017/xxxxx}

\aopheadtitle{Proceedings IAU Symposium}
\editors{C. Sterken,  J. Hearnshaw \&  D. Valls-Gabaud, eds.}

\title{What drives kpc-scale outflows in Radio-Quiet AGN? Insights from a Polarimetric Study}

\author{Salmoli Ghosh$^{1}$, Preeti Kharb$^{1}$, Biny Sebastian$^{2,3}$, Jack Gallimore$^{4}$, Alice Pasetto$^{5}$, Christopher P. O'Dea$^{2}$, Timothy Heckman$^{6}$, Stefi A. Baum$^{2}$}
\affiliation{$^{1}$National Centre for Radio Astrophysics (NCRA) - Tata Institute of Fundamental Research (TIFR), S. P. Pune University Campus, Pune 411007, Maharashtra, India\\
$^{2}$Department of Physics and Astronomy, University of Manitoba, Winnipeg, Manitoba, Canada\\
$^{3}$Space Telescope Science Institute, 3700 San Martin Drive, Baltimore, MD 21218, USA\\
$^{4}$Department of Physics and Astronomy, Bucknell University, Lewisburg, PA 17837\\
$^{5}$Instituto de Radioastronom\'ia y Astrof\'isica (IRyA-UNAM), 3-72 (Xangari), 8701, Morelia, Mexico\\
$^{6}$William H. Miller III Department of Physics \& Astronomy, Johns Hopkins University, Bloomberg Center for Physics and Astronomy, 3400 N. Charles Street, Baltimore, MD 21218, USA}

\begin{abstract}
We present a review of our findings on the origin, drivers, nature, and impact of kiloparsec-scale radio emission in radio-quiet (RQ) AGN. Using radio polarimetric techniques, we probe the dynamics and magnetic (B-) field geometry of outflows in Seyfert and LINER galaxies. Multi-band data from the Karl G. Jansky Very Large Array (VLA) reveal how low-power jets interact with their environment. These interactions can slow down and disrupt the radio outflows while locally regulating star formation through AGN feedback. Several radio properties correlate strongly with the black hole mass, similar to trends observed in radio-loud (RL) AGN. Although their characteristics differ, RQ systems might not be intrinsically distinct from RL AGN, apart from lower jet powers. Our polarization measurements further suggest a composite model in which a black hole–accretion disk system drives both a collimated jet with a small-pitch-angle helical B-field and a wide-angle wind threaded by a high-pitch-angle helical field.
\end{abstract}


\maketitle

\section{Introduction}
Active Galactic Nuclei (AGN) are luminous and energetic centers of massive galaxies, powered by the accretion of matter onto supermassive black holes \citep[SMBHs;][]{Rees1984}. 
Many AGN launch relativistic jets that are prominent at radio frequencies. While the majority of AGN ($\sim$80\%) are radio-quiet \citep[RQ;][]{Kellermann1989}, their weak radio emission and the absence of powerful, large-scale jets have left them comparatively less explored. Long-standing open questions concerning RQ AGN include:
(i) What is the origin of their radio emission?
(ii) Why are their jets confined within host galaxies rather than extending beyond?
(iii) Do their outflows exhibit episodic or stratified structures?
(iv) What mechanisms drive these outflows?
(v) What is the nature and impact of AGN feedback in RQ AGN? 
The radio emission in RQ AGN has been attributed to several possible origins, including coronal emission, stellar processes, shocks associated with emission-line gas that drive galactic synchrotron winds, or AGN-driven outflows in the form of collimated jets or wide-angle winds \citep{Laor2008, Harrison2014, Panessa2019, Sebastian2019, Kharb2023}. 
Outflows launched by AGN are expected to exhibit significant polarization due to the presence of ordered magnetic (B-) fields. Polarization measurements provide a means of tracing magnetic-field orientations in the plane of the sky and offer key insights into the underlying nature of the outflows. Multi-frequency polarization observations provide constraints on the properties of the surrounding medium through the measurement of Faraday rotation.

To address the above questions, we selected a subsample of 26 Seyfert or Low-Ionization Nuclear Emission-line Region (LINER) galaxies from the Centre for Astrophysics (CfA) + extended 12 $\mu$m sample, as compiled by \citet{Gallimore2006}. These authors reported that more than $44\%$ of Seyferts/LINERs exhibit radio structures extending up to a few kiloparsecs, referred to as kiloparsec-scale radio structures (KSRs). 
We observed 12 such KSRs with the Karl G. Jansky Very Large Array (VLA), at 10 GHz in the D-configuration and at 1.4 GHz in BnA$\rightarrow$A configuration, achieving angular resolutions of 7 arcsec and 2 arcsec, respectively \citep{Ghosh2025}. 
Data analysis for the remaining sources in the sample is ongoing and will be presented in a forthcoming paper. This sample is well-studied across multiple wavelengths, enabling a comprehensive assessment of the diverse physical processes at work.

\section{Polarimetric Observations of Seyfert and LINER galaxies with KSRs}
Polarized radio emission was detected in all sources of the sample. In four nearby, star-forming galaxies, viz., NGC 4388, NGC 3079, NGC 4594, and NGC 1068, polarized emission was also observed from their galactic disks. In contrast, AGN-driven radio outflows were predominantly extraplanar, extending along the host galaxy’s minor axes, and exhibited a systematically higher polarization degree than the galactic disk emission. While many sources showed no detectable polarization in the cores, polarized emission was consistently observed in the extended regions, reaching values as high as 47\% in the lobes of NGC 1320 at 1.4 GHz. The inferred B-fields, oriented perpendicular to the observed polarization vectors in optically thin regions, revealed bent or disrupted jets that were not easily evident in total-intensity images. In most Seyfert or LINER galaxy lobes, the B-fields are aligned with the lobe axes, consistent with confinement by the surrounding medium \citep{Laing1980a,Ghosh2023}. In many radio cores, the B-fields were oriented perpendicular to the jet axes, suggestive of toroidal fields associated with nuclear winds \citep[][]{Mehdipour2019}. NGC 3516 showed perpendicular B-fields in the core, aligned B-fields with the radio outflow away from the core, and perpendicular B-fields in its terminal shocked region. The resulting B-field structure closely resembled that of radio-loud (RL) Fanaroff Riley type II radio galaxies \citep[e.g.,][]{Kharb2008, Perley2017, Baghel2024}. Thus, RQ AGN outflows can also exhibit high degrees of polarization and ordered B-fields, terminating in shocks that are similar to the morphology of their RL counterparts, albeit on smaller spatial scales. 

\section{Interactions between Multiphase Gas \& Radio Outflows}
We found evidence for jet–medium interactions in several Seyfert and LINER galaxies in our sample, identified through multi-band data in combination with radio data. A few representative cases are now discussed. NGC 3079 is an edge-on Seyfert+starburst galaxy in which the parsec- and kpc-scale jets are misaligned \citep{Irwin2003}. The inclined jet appears to interact with material from the surrounding galactic disk and is subsequently redirected along the galaxy’s minor axis \citep{Irwin1988, Mukherjee2018}. The HI image from the Continuum Halos in Nearby Galaxies - an EVLA Survey \citep[CHANG-ES;][]{Irwin2012}
reveals a deficit of neutral gas along the Seyfert outflow, likely due to AGN-driven ionization. Co-spatial H$\alpha$ emission supports this interpretation, suggesting that the reduced HI reservoir suppresses star formation, consistent with negative AGN feedback.

In NGC 2992, the jet orientation traced by the VLA and Very Long Baseline Interferometry (VLBI) differs, indicating a change in jet direction. \citet{Irwin2017} reported relic emission in this source, coincident with the highly polarized region identified in our study. This region overlaps with one arm of the H$\alpha$ emission from \citep{Vargas2019}, suggesting a compression of gas and B-fields in the relic lobe by subsequent episodes of jet activity. A similar case is observed in NGC 4388, where the jet bends toward regions of enhanced H$\alpha$ emission \citep[from][]{Vargas2019} along the galactic spiral arm. The deficit of atomic hydrogen near the jet may result from enhanced star formation, consistent with positive AGN feedback. 

NGC 3516 exhibits an S-shaped radio morphology, which is also seen in the [O III] 5007$\AA$ emission but on smaller spatial scales. This structure may arise from a precessing jet entraining narrow-line region gas, or from ram-pressure bending induced by galactic rotation \citep{Veilleux1993}. Chandra X-ray images reveal enhanced emission along the brighter limb of the outflow \citep[see][]{Ghosh2025b}. The radio outflow deviates as it approaches the shock front, consistent with either a disrupted jet or a wide-angle, disk-driven outflow. 

Taken together, these results reveal the coexistence of multiphase gas along the AGN-driven synchrotron outflows in all sources with multi-band data. The multiphase gas interacts with the radio plasma, acting as a Faraday rotating medium that partly depolarize the emission and slow down the outflow. In turn, the nuclear radio outflow ionizes, entrains, evacuates, and compresses the surrounding medium, either enhancing or suppressing star formation in the host galaxy on local scales. We therefore conclude that RQ AGN outflows are capable of driving both negative and positive feedback on small spatial scales.

\section{Driver of Radio Outflows in RQ AGN}
We found that most of the radio properties measured at 10 GHz and 1.4 GHz, like radio loudness \citep[see][]{Melendez2010}, core luminosity, total energy from equipartition estimates \citep[following relations in][]{OdeaOwen1987}, and jet kinetic power \citep[following][]{Merloni2007, Foschini2014}, show strong to moderate positive correlations with SMBH masses. Similar tests using the Eddington ratio as a proxy for accretion rate, combined with partial correlation analysis to disentangle the effects of mass and accretion, indicated that the dependence on black hole mass strengthens, while that on accretion rate weakens. Although the sample size is small (12 sources), these results suggest that black hole mass plays a dominant role in driving RQ outflows in KSRs. All sources in the sample are low-accretion systems. We found an anti-correlation between the spectral index and accretion rate or Eddington ratio, consistent with the results from \citet{Laor2019}.
This trend suggests that highly accreting systems generate steeper, more extended outflows.
A similar dependence is observed between fractional polarization and accretion rate. In the cores, the fractional polarization anti-correlates with accretion rate, implying that enhanced accretion brings in more matter near the nucleus, thereby increasing depolarization. In contrast, in the extended outflows, the fractional polarization correlates positively with accretion rate, indicating that higher accretion promotes the development of strongly polarized, powerful outflows dominated by ordered B-fields.

\section{Nature of the Radio Outflows -- Insights from Polarimetric Observations}
RQ sources such as NGC 2992 exhibit relic emission detected in polarization but not in total intensity \citep{Irwin2017}. Other RQ sources, including Mrk~6, NGC 2639 and NGC 3516, show evidence for multiple episodes of activity \citep{Kharb2006, rao2023, Ghosh2025b}, with the oldest emission extending over a few kpc. 
Jet–medium interactions may play a critical role in decelerating and disrupting RQ AGN-driven radio outflows. This is supported by our multi-wavelength analysis of Seyferts and LINERs, as well as depolarization measurements that indicate plasma mixing. Although the inferred electron densities of this mixed plasma are low ($10^{-3}$–$10^{-1}$ cm$^{-3}$), the observed effects can be explained by the presence of thin, filamentary layers of gas, or a sheath of magnetized wind.

\begin{figure}
    \centering
    \includegraphics[width=13cm,trim=30 100 30 80]{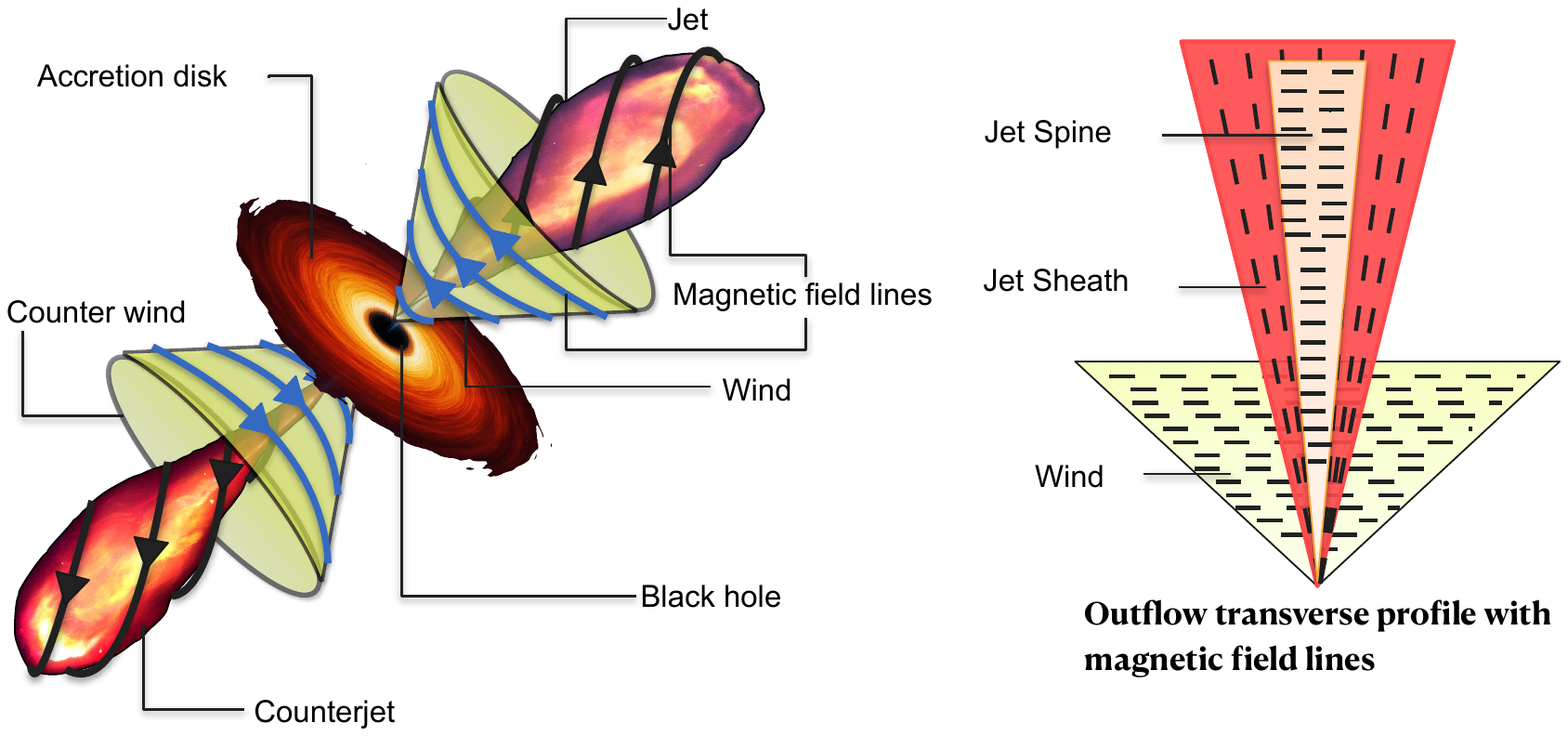}
    \caption{Left: A proposed model for radio outflows in RQ AGN with KSRs, based on radio polarimetric observations. Right: The projected magnetic field orientations in the different layers of the RQ outflows.}
    \label{fig:jetwind}
\end{figure}

The coexistence of aligned and perpendicular B-fields indicates the presence of both poloidal and toroidal B-field components, forming helical field structures. Our detection of transverse rotation measure gradients and circular polarization \citep[e.g.,][]{Ghosh2025b} in several sources further supports this scenario. The toroidal component is consistent with a disk-driven, wide-angle winds or outflows characterized by helices with large pitch angles, while collimated jet outflows dominated by poloidal fields likely correspond to helices with smaller pitch angles \citep[also see][]{Gabuzda2008}. In projection, stretched helices appear as aligned fields, whereas compressed helices are observed as perpendicular fields. Evidence for such complexity is also found in NGC 4151, where we detect a stratified jet (Ghosh et al. in preparation). The magnetic field is perpendicular in the “jet spine,” tracing shocked emission along the axis; parallel in the “jet sheath,” shaped by shear at the jet–medium interface; and again perpendicular in the outer “wind” region. A schematic of this proposed radio outflow structure in RQ AGN with KSRs is shown in Figure \ref{fig:jetwind}. A similar spine–sheath structure has been reported in RL sources \citep{Attridge1999, Laing2006}. Our observations suggest that both stratified jets and winds are present in RQ sources with KSRs. 

\bibliography{iauguide}{}

\begin{thebibliography}{}
\makeatletter
\relax
\def\mn@urlcharsother{\let\do\@makeother \do\$\do\&\do\#\do\^\do\_\do\%\do\~}
\def\mn@doi{\begingroup\mn@urlcharsother \@ifnextchar [ {\mn@doi@} {\mn@doi@[]}}
\def\mn@doi@[#1]#2{\def\@tempa{#1}\ifx\@tempa\@empty \href {http://dx.doi.org/#2} {doi:#2}\else \href {http://dx.doi.org/#2} {#1}\fi \endgroup}
\def\mn@eprint#1#2{\mn@eprint@#1:#2::\@nil}
\def\mn@eprint@arXiv#1{\href {http://arxiv.org/abs/#1} {{\tt arXiv:#1}}}
\def\mn@eprint@dblp#1{\href {http://dblp.uni-trier.de/rec/bibtex/#1.xml} {dblp:#1}}
\def\mn@eprint@#1:#2:#3:#4\@nil{\def\@tempa {#1}\def\@tempb {#2}\def\@tempc {#3}\ifx \@tempc \@empty \let \@tempc \@tempb \let \@tempb \@tempa \fi \ifx \@tempb \@empty \def\@tempb {arXiv}\fi \@ifundefined {mn@eprint@\@tempb}{\@tempb:\@tempc}{\expandafter \expandafter \csname mn@eprint@\@tempb\endcsname \expandafter{\@tempc}}}

\bibitem[\protect\citeauthoryear{{Attridge}, {Roberts}  \& {Wardle}}{{Attridge} et~al.}{1999}]{Attridge1999}
{Attridge} J.~M.,  {Roberts} D.~H.,   {Wardle} J. F.~C.,  1999, \mn@doi [\apjl] {10.1086/312078}, \href {https://ui.adsabs.harvard.edu/abs/1999ApJ...518L..87A} {518, L87}

\bibitem[\protect\citeauthoryear{{Baghel}, {Kharb}, {Hovatta}, {Ho}, {Harrison}, {Lindfors}, {Silpa}  \& {Gulati}}{{Baghel} et~al.}{2024}]{Baghel2024}
{Baghel} J.,  {Kharb} P.,  {Hovatta} T.,  {Ho} L.~C.,  {Harrison} C.,  {Lindfors} E.,  {Silpa} S.,   {Gulati} S.,  2024, \mn@doi [\apj] {10.3847/1538-4357/ad8d58}, \href {https://ui.adsabs.harvard.edu/abs/2024ApJ...977..192B} {977, 192}

\bibitem[\protect\citeauthoryear{{Foschini}}{{Foschini}}{2014}]{Foschini2014}
{Foschini} L.,  2014, in International Journal of Modern Physics Conference Series. p. 1460188 (\mn@eprint {arXiv} {1310.5822}), \mn@doi{10.1142/S2010194514601884}

\bibitem[\protect\citeauthoryear{{Gabuzda}, {Vitrishchak}, {Mahmud}  \& {O'Sullivan}}{{Gabuzda} et~al.}{2008}]{Gabuzda2008}
{Gabuzda} D.~C.,  {Vitrishchak} V.~M.,  {Mahmud} M.,   {O'Sullivan} S.~P.,  2008, \mn@doi [\mnras] {10.1111/j.1365-2966.2007.12773.x}, \href {https://ui.adsabs.harvard.edu/abs/2008MNRAS.384.1003G} {384, 1003}

\bibitem[\protect\citeauthoryear{{Gallimore}, {Axon}, {O'Dea}, {Baum}  \& {Pedlar}}{{Gallimore} et~al.}{2006}]{Gallimore2006}
{Gallimore} J.~F.,  {Axon} D.~J.,  {O'Dea} C.~P.,  {Baum} S.~A.,   {Pedlar} A.,  2006, \mn@doi [\aj] {10.1086/504593}, \href {http://adsabs.harvard.edu/abs/2006AJ....132..546G} {132, 546}

\bibitem[\protect\citeauthoryear{{Ghosh}, {Kharb}, {Baghel}  \& {Silpa}}{{Ghosh} et~al.}{2023}]{Ghosh2023}
{Ghosh} S.,  {Kharb} P.,  {Baghel} J.,   {Silpa} S.,  2023, \mn@doi [\apj] {10.3847/1538-4357/acfa00}, \href {https://ui.adsabs.harvard.edu/abs/2023ApJ...958...71G} {958, 71}

\bibitem[\protect\citeauthoryear{{Ghosh}, {Kharb}, {Sebastian}, {Gallimore}, {Pasetto}, {O'Dea}, {Heckman}  \& {Baum}}{{Ghosh} et~al.}{2025a}]{Ghosh2025}
{Ghosh} S.,  {Kharb} P.,  {Sebastian} B.,  {Gallimore} J.,  {Pasetto} A.,  {O'Dea} C.~P.,  {Heckman} T.,   {Baum} S.~A.,  2025a, \mn@doi [\apj] {10.3847/1538-4357/adae05}, \href {https://ui.adsabs.harvard.edu/abs/2025ApJ...982..141G} {982, 141}

\bibitem[\protect\citeauthoryear{{Ghosh}, {Kharb}, {Sajjanhar}, {Pasetto}  \& {Sebastian}}{{Ghosh} et~al.}{2025b}]{Ghosh2025b}
{Ghosh} S.,  {Kharb} P.,  {Sajjanhar} E.,  {Pasetto} A.,   {Sebastian} B.,  2025b, \mn@doi [\apj] {10.3847/1538-4357/ade98d}, \href {https://ui.adsabs.harvard.edu/abs/2025ApJ...989...40G} {989, 40}

\bibitem[\protect\citeauthoryear{{Harrison}, {Alexander}, {Mullaney}  \& {Swinbank}}{{Harrison} et~al.}{2014}]{Harrison2014}
{Harrison} C.~M.,  {Alexander} D.~M.,  {Mullaney} J.~R.,   {Swinbank} A.~M.,  2014, \mn@doi [\mnras] {10.1093/mnras/stu515}, \href {https://ui.adsabs.harvard.edu/abs/2014MNRAS.441.3306H} {441, 3306}

\bibitem[\protect\citeauthoryear{{Irwin} \& {Saikia}}{{Irwin} \& {Saikia}}{2003}]{Irwin2003}
{Irwin} J.~A.,  {Saikia} D.~J.,  2003, \mn@doi [\mnras] {10.1111/j.1365-2966.2003.07146.x}, \href {https://ui.adsabs.harvard.edu/abs/2003MNRAS.346..977I} {346, 977}

\bibitem[\protect\citeauthoryear{{Irwin} \& {Seaquist}}{{Irwin} \& {Seaquist}}{1988}]{Irwin1988}
{Irwin} J.~A.,  {Seaquist} E.~R.,  1988, \mn@doi [\apj] {10.1086/166956}, \href {https://ui.adsabs.harvard.edu/abs/1988ApJ...335..658I} {335, 658}

\bibitem[\protect\citeauthoryear{{Irwin} et~al.,}{{Irwin} et~al.}{2012}]{Irwin2012}
{Irwin} J.,  et~al., 2012, \mn@doi [\aj] {10.1088/0004-6256/144/2/43}, \href {https://ui.adsabs.harvard.edu/abs/2012AJ....144...43I} {144, 43}

\bibitem[\protect\citeauthoryear{{Irwin} et~al.,}{{Irwin} et~al.}{2017}]{Irwin2017}
{Irwin} J.~A.,  et~al., 2017, \mn@doi [\mnras] {10.1093/mnras/stw2414}, \href {https://ui.adsabs.harvard.edu/abs/2017MNRAS.464.1333I} {464, 1333}

\bibitem[\protect\citeauthoryear{{Kellermann}, {Sramek}, {Schmidt}, {Shaffer}  \& {Green}}{{Kellermann} et~al.}{1989}]{Kellermann1989}
{Kellermann} K.~I.,  {Sramek} R.,  {Schmidt} M.,  {Shaffer} D.~B.,   {Green} R.,  1989, \mn@doi [\aj] {10.1086/115207}, \href {https://ui.adsabs.harvard.edu/abs/1989AJ.....98.1195K} {98, 1195}

\bibitem[\protect\citeauthoryear{{Kharb} \& {Silpa}}{{Kharb} \& {Silpa}}{2023}]{Kharb2023}
{Kharb} P.,  {Silpa} S.,  2023, \mn@doi [Galaxies] {10.3390/galaxies11010027}, \href {https://ui.adsabs.harvard.edu/abs/2023Galax..11...27K} {11, 27}

\bibitem[\protect\citeauthoryear{{Kharb}, {O'Dea}, {Baum}, {Colbert}  \& {Xu}}{{Kharb} et~al.}{2006}]{Kharb2006}
{Kharb} P.,  {O'Dea} C.~P.,  {Baum} S.~A.,  {Colbert} E.~J.~M.,   {Xu} C.,  2006, \mn@doi [\apj] {10.1086/507945}, \href {https://ui.adsabs.harvard.edu/abs/2006ApJ...652..177K} {652, 177}

\bibitem[\protect\citeauthoryear{{Kharb}, {O'Dea}, {Baum}, {Daly}, {Mory}, {Donahue}  \& {Guerra}}{{Kharb} et~al.}{2008}]{Kharb2008}
{Kharb} P.,  {O'Dea} C.~P.,  {Baum} S.~A.,  {Daly} R.~A.,  {Mory} M.~P.,  {Donahue} M.,   {Guerra} E.~J.,  2008, \mn@doi [\apjs] {10.1086/520840}, \href {http://adsabs.harvard.edu/abs/2008ApJS..174...74K} {174, 74}

\bibitem[\protect\citeauthoryear{{Laing}}{{Laing}}{1980}]{Laing1980a}
{Laing} R.~A.,  1980, in Bulletin of the American Astronomical Society. p.~823

\bibitem[\protect\citeauthoryear{{Laing}, {Canvin}, {Cotton}  \& {Bridle}}{{Laing} et~al.}{2006}]{Laing2006}
{Laing} R.~A.,  {Canvin} J.~R.,  {Cotton} W.~D.,   {Bridle} A.~H.,  2006, \mn@doi [\mnras] {10.1111/j.1365-2966.2006.10099.x}, \href {https://ui.adsabs.harvard.edu/abs/2006MNRAS.368...48L} {368, 48}

\bibitem[\protect\citeauthoryear{{Laor} \& {Behar}}{{Laor} \& {Behar}}{2008}]{Laor2008}
{Laor} A.,  {Behar} E.,  2008, \mn@doi [\mnras] {10.1111/j.1365-2966.2008.13806.x}, \href {http://adsabs.harvard.edu/abs/2008MNRAS.390..847L} {390, 847}

\bibitem[\protect\citeauthoryear{{Laor}, {Baldi}  \& {Behar}}{{Laor} et~al.}{2019}]{Laor2019}
{Laor} A.,  {Baldi} R.~D.,   {Behar} E.,  2019, \mn@doi [\mnras] {10.1093/mnras/sty3098}, \href {https://ui.adsabs.harvard.edu/abs/2019MNRAS.482.5513L} {482, 5513}

\bibitem[\protect\citeauthoryear{{Mehdipour} \& {Costantini}}{{Mehdipour} \& {Costantini}}{2019}]{Mehdipour2019}
{Mehdipour} M.,  {Costantini} E.,  2019, \mn@doi [\aap] {10.1051/0004-6361/201935205}, \href {https://ui.adsabs.harvard.edu/abs/2019A&A...625A..25M} {625, A25}

\bibitem[\protect\citeauthoryear{{Mel{\'e}ndez}, {Kraemer}  \& {Schmitt}}{{Mel{\'e}ndez} et~al.}{2010}]{Melendez2010}
{Mel{\'e}ndez} M.,  {Kraemer} S.~B.,   {Schmitt} H.~R.,  2010, \mn@doi [\mnras] {10.1111/j.1365-2966.2010.16679.x}, \href {https://ui.adsabs.harvard.edu/abs/2010MNRAS.406..493M} {406, 493}

\bibitem[\protect\citeauthoryear{{Merloni} \& {Heinz}}{{Merloni} \& {Heinz}}{2007}]{Merloni2007}
{Merloni} A.,  {Heinz} S.,  2007, \mn@doi [\mnras] {10.1111/j.1365-2966.2007.12253.x}, \href {http://adsabs.harvard.edu/abs/2007MNRAS.381..589M} {381, 589}

\bibitem[\protect\citeauthoryear{{Mukherjee}, {Bicknell}, {Wagner}, {Sutherland}  \& {Silk}}{{Mukherjee} et~al.}{2018}]{Mukherjee2018}
{Mukherjee} D.,  {Bicknell} G.~V.,  {Wagner} A.~Y.,  {Sutherland} R.~S.,   {Silk} J.,  2018, \mn@doi [\mnras] {10.1093/mnras/sty1776}, \href {https://ui.adsabs.harvard.edu/abs/2018MNRAS.479.5544M} {479, 5544}

\bibitem[\protect\citeauthoryear{{O'Dea} \& {Owen}}{{O'Dea} \& {Owen}}{1987}]{OdeaOwen1987}
{O'Dea} C.~P.,  {Owen} F.~N.,  1987, \mn@doi [\apj] {10.1086/165182}, \href {http://adsabs.harvard.edu/cgi-bin/nph-bib_query?bibcode=1987ApJ...316...95O&db_key=AST} {316, 95}

\bibitem[\protect\citeauthoryear{{Panessa}, {Baldi}, {Laor}, {Padovani}, {Behar}  \& {McHardy}}{{Panessa} et~al.}{2019}]{Panessa2019}
{Panessa} F.,  {Baldi} R.~D.,  {Laor} A.,  {Padovani} P.,  {Behar} E.,   {McHardy} I.,  2019, \mn@doi [Nature Astronomy] {10.1038/s41550-019-0765-4}, \href {https://ui.adsabs.harvard.edu/abs/2019NatAs...3..387P} {3, 387}

\bibitem[\protect\citeauthoryear{{Perley} \& {Meisenheimer}}{{Perley} \& {Meisenheimer}}{2017}]{Perley2017}
{Perley} R.~A.,  {Meisenheimer} K.,  2017, \mn@doi [\aap] {10.1051/0004-6361/201629704}, \href {https://ui.adsabs.harvard.edu/abs/2017A&A...601A..35P} {601, A35}

\bibitem[\protect\citeauthoryear{{Rao} et~al.,}{{Rao} et~al.}{2023}]{rao2023}
{Rao} V.~V.,  et~al., 2023, \mn@doi [\mnras] {10.1093/mnras/stad1901}, \href {https://ui.adsabs.harvard.edu/abs/2023MNRAS.524.1615R} {524, 1615}

\bibitem[\protect\citeauthoryear{{Rees}}{{Rees}}{1984}]{Rees1984}
{Rees} M.~J.,  1984, \mn@doi [\araa] {10.1146/annurev.aa.22.090184.002351}, \href {https://ui.adsabs.harvard.edu/abs/1984ARA&A..22..471R} {22, 471}

\bibitem[\protect\citeauthoryear{{Sebastian}, {Kharb}, {O' Dea}, {Gallimore}, {Baum}  \& {Colbert}}{{Sebastian} et~al.}{2019}]{Sebastian2019}
{Sebastian} B.,  {Kharb} P.,  {O' Dea} C.~P.,  {Gallimore} J.~F.,  {Baum} S.~A.,   {Colbert} E.~J.~M.,  2019, \mn@doi [arXiv e-prints] {10.48550/arXiv.1912.09511}, \href {https://ui.adsabs.harvard.edu/abs/2019arXiv191209511S} {p. arXiv:1912.09511}

\bibitem[\protect\citeauthoryear{{Vargas}, {Walterbos}, {Rand}, {Stil}, {Krause}, {Li}, {Irwin}  \& {Dettmar}}{{Vargas} et~al.}{2019}]{Vargas2019}
{Vargas} C.~J.,  {Walterbos} R. A.~M.,  {Rand} R.~J.,  {Stil} J.,  {Krause} M.,  {Li} J.-T.,  {Irwin} J.,   {Dettmar} R.-J.,  2019, \mn@doi [\apj] {10.3847/1538-4357/ab27cb}, \href {https://ui.adsabs.harvard.edu/abs/2019ApJ...881...26V} {881, 26}

\bibitem[\protect\citeauthoryear{{Veilleux}, {Tully}  \& {Bland-Hawthorn}}{{Veilleux} et~al.}{1993}]{Veilleux1993}
{Veilleux} S.,  {Tully} R.~B.,   {Bland-Hawthorn} J.,  1993, \mn@doi [\aj] {10.1086/116512}, \href {https://ui.adsabs.harvard.edu/abs/1993AJ....105.1318V} {105, 1318}

\makeatother
\end{thebibliography}
\bibliographystyle{mnras}
\end{document}